# Size-dependent radiosensitization of PEG-coated gold nanoparticles for cancer radiation therapy

*Xiao-Dong Zhang [a,\*], Di Wu [a], Xiu Shen [a], Jie Chen [a], Yuan-Ming Sun [a], Pei-Xun Liu [a],*

*Xing-Jie Liang [b,\*]*

[a] Institute of Radiation Medicine and Tianjin Key Laboratory of Molecular Nuclear Medicine, Chinese Academy of Medical Sciences and Peking Union Medical College, Tianjin 300192, People's Republic of China.

[b] Laboratory of Nanomedicine and Nanosafety, Division of Nanomedicine and Nanobiology, National Center for Nanoscience and Technology, China, and CAS Key Laboratory for Biomedical Effects of Nanomaterials and Nanosafety, Chinese Academy of Sciences, Beijing 100190, People's Republic of China.

**ABSTRACT**: Gold nanoparticles have been conceived as a radiosensitizer in cancer radiation therapy, but one of the important questions for primary drug screening is what size of gold nanoparticles can optimally enhance radiation effects. Herein, we perform *in vitro* and *in vivo* radiosensitization studies of 4.8, 12.1, 27.3, and 46.6 nm PEG-coated gold nanoparticles. *In vitro* results show that all sizes of the PEG-coated gold nanoparticles can cause a significant decrease in cancer cell survival after gamma radiation. 12.1 and 27.3 nm PEG-coated gold nanoparticles have dispersive distributions in the cells and have stronger sensitization effects than 4.8 and 46.6 nm particles by both cell apoptosis and necrosis. Further, *in vivo* results also show all sizes of the PEG-coated gold nanoparticles can decrease tumor volume and weight after 5 Gy radiations, and 12.1 and 27.3 nm PEG-coated gold nanoparticles have greater sensitization effects than 4.8 and 46.6 nm particles, which can lead to almost complete disappearance of the tumor. *In vivo* biodistribution confirms that 12.1 and 27.3 nm PEG-coated gold nanoparticles are accumulated in the tumor with high concentrations. The pathology, immune response, and blood biochemistry indicate that the PEG-coated gold nanoparticles don't cause spleen and kidney damages, but give rise to liver damage and gold accumulation. It can be concluded that 12.1 and 27.3 nm PEG-coated gold nanoparticles show high radiosensitivity, and these results have an important indication for possible radiotherapy and drug delivery.

---

[\*] Author to whom correspondence should be addressed. Electronic-mail:

Xiaodongzhang@irm.cams.ac. cn.

[\*] Author to whom correspondence should be addressed. Electronic-mail: liangxj@nanoctr.cn.

# 1. Introduction

Gold nanoparticles (NPs) have wide biomedical applications because of their unique surface chemistry, electronic, and optical properties [1-2]. Today, gold NPs have been conceived as a kind of radiosensitizer in radiotherapy. This is because the strong photoelectric absorption and secondary electron caused by gamma or X-ray irradiation can accelerate DNA strand break [3-4]. The biosafety of metallic gold is good and it has been used since the 1950s. *In vitro* and *in vivo* biocompatibilities of gold NPs have been widely investigated in recent years [5-6]. The toxicity of gold NPs can be determined by size, shape, surface charge, and surface coating, but the overall toxicity dose of gold NPs is in an acceptable level [7-12]. This makes the applications of gold NPs in cancer radiation therapy possible.

In general, the photoelectric absorption cross sections of radiosensitive materials directly depend on atom radius [13], and thus their radiotherapy effects also depend on atom radius [14-17]. This physical principle determines that gold (atomic-number Z=79) has a stronger radiation enhancement effect by photoelectric effect than other radiosensitive materials such as carbon (Z=6), iodine (Z = 53), gadolinium (Z = 64), and platinum (Z =78). Gold NPs have a larger atomic radius than gold atom (~0.17 nm), and thus should have a stronger radiation enhancement effect than gold atom. Further, in theory, the enhanced effect of gold NPs in radiotherapy should be size-dependent, namely the larger, the better. However, the real situation is complex.

*In vivo* experimental data showed that small gold NPs of 1.9 nm could effectively enhance radiotherapy [3, 18]. Subsequently, *in vitro* investigation showed that the PEG-coated gold NPs of 4.6 and 6.1 nm could decrease the cell survival rates of both EMT-6 and CT26 [19]. Typical 10 nm gold NPs coated with cysteamine and glucose had a significant radiation sensitivity to all high energy rays, with 200 keV being preferred [20]. However, it was surprising that gold NPs in the size range 5-10 nm were not found to be more efficient in enhancement of radiation sensitivity as compared with the ultra small 1.9 nm gold particles [21]. In other work, it was found that 50 nm naked gold NPs have a stronger radiation sensitivity than 14 and 74 nm gold particles. This was because 50 nm naked gold particles had the greatest endocytosis and cellar uptake [4, 22]. However, it is well known that physiological pH citrate coated gold NPs have a high zeta potential, and the particles tend to aggregate severely [23]. Therapeutic gold NPs smaller than 50 nm can easily pass through cell membrane, and particles smaller than 20 nm can pass through blood vessel endothelium [24]. Meanwhile, the surface coating, such as PEG-SH, can improve the stability of gold NPs [25-26]. Therefore, it is desirable to clarify the size-dependent radiosensitization effects of 5-60 nm PEG-coated gold NPs. This has a significant indication on the effectiveness of both radiotherapy and drug delivery.

Here, we address *in vitro* and *in vivo* size-dependent enhancement of radiation effects of gold NPs. PEG-SH, widely used as coating material, was chosen as the surface coating to improve the monodispersity and biocompatibility of gold NPs. We carried out the radiosensitization studies of 4.8, 12.1, 27.3, and 46.6 nm PEG-coated gold NPs in HeLa cells. The toxicities of the PEG-coated gold NPs with different sizes were analyzed after 24 and 48 hours treatments. Cloning formation and cell apoptosis were used to evaluate the enhancement of radiation effects. Subsequently, *in vivo* radiosensitization effects of 4.8, 12.1, 27.3, and 46.6 nm PEG-coated gold particles were studied by evaluating tumor size and weight after 5 Gy gamma radiations. The pathology, biochemistry, organ index, and distribution were used to evaluate the toxicities of these PEG-coated gold NPs. These experimental results will clarify what size of gold particles is suitable for radiotherapy and related medical treatments.

# 2. Materials and methods

## 2.1. Fabrication and stability of the PEG-coated gold NPs

Gold NPs were fabricated following the classical method introduced by Turkevich. 100 mL 0.01% chlorauric acid ($HAuCl_4 \cdot 4H_2O$) solution was refluxed, and 0.8, 1.3, and 5 mL 1% sodium citrate solution was added respectively to the boiling solution for obtaining NPs with different sizes. The reduction of gold ions by the citrate ions was completed after 5 min. The solution was further boiled for 30 min, and then left to cool at room temperature. The small size gold NPs were reduced by $NaBH_4$. Subsequently, 1 mg PEG-SH (5000 MW, Sigma) was mixed with gold NPs and stirred for 1 h to covalently modify the surface of the gold NPs with PEG[2]. The resulting PEG-coated gold NPs were collected by centrifugation at 16,000 rpm for 30 min and washed twice with distilled water. The PEG-coated gold NPs solution was stored at 4 ℃ in order to prevent aggravations.

Zeta (ζ) potential for gold NPs was determined with the NanoZS Zetasizer particle analyzer (Malvern Instruments, Ltd., Malvern, UK). Data were acquired in the phase analysis light scattering mode at 25 °C, and sample solutions were prepared by diluting gold NPs into 10 mM phosphate-buffered saline (PBS) solution (pH 7.0). Stability of gold NPs was evaluated using optical absorption. Gold NPs were diluted five times in human blood plasma, and optical absorption spectra were measured at the time points of 0.5, 1, 2, 6, 12, and 24 hours using a UV-vis spectrophotometer (DU800) in a 5 ml glass cuvette. The size and morphology of gold NPs were analyzed by transmission electron microscopy (TEM) using a Hitachi HF-2000 field emission high-resolution TEM operating at 200 kV.

*2.2. Cell culture and viability assay*

HeLa cells were cultured in low-glucose DMEM. Media contained fetal calf serum (10 %), L-glutamine (2.9 mg/mL), streptomycin (1 mg/mL), and penicillin (1000 units/mL) at 37 °C in humidified atmosphere with 5% $CO_2$. The cells (in culture medium) were dispensed in 96-well plates (90 mL in each well containing $10^4$ cells). 10 μL PEG-coated gold NPs were dissolved in culture medium, and then 100 μL blendings were added to each well with different concentrations (0.001-025 mM). The concentration effect of the PEG-coated gold NPs was assessed using Cell Titre-Glo™ luminescent cell viability assay (Promega, Madison, WI, USA) after 24 and 48 hours treatments. This assay was a homogenous method of determining the number of viable cells in culture based on the quantitation of adenosine triphosphate (ATP), which signaled the presence of metabolically active cells. After the treatment, the cells were incubated with 20 μL of Cell Titre-Glo™ reagent and contents were allowed to mix on an orbital shaker in accordance with the assay protocols, which resulted in cell lysis and generation of a luminescent signal proportional to the amount of ATP present. The amount of ATP was proportional to the number of cells presented in culture. The luminescence signal was recorded with a single tube luminometer (TD 20/20, Turner Biosystems Inc., Sunnyvale, CA, USA).

A four-parameter logistic model was used to evaluate half maximal inhibitory concentration ($IC_{50}$) of gold NPs. For each size of the PEG-coated gold NPs, $IC_{50}$ value was determined from triplicate wells during both the stationary and logarithmic cell growth phases. $IC_{50}$ value obtained from logarithmic cell growth was routinely repeated in three independent experiments.

*2.3. Colony formation assay*

HeLa cells were incubated in 25 $cm^2$ flasks overnight and then exposed to the PEG-coated gold NPs at the concentrations of 0.05 and 0.1 mM. Gold NPs were administrated 24 hours before irradiation. The cells were irradiated by $^{137}Cs$ with activity of 3600 Ci and photon energy of 662 KeV at doses of 1, 2, 4, 6, and 8 Gy, respectively. After irradiation, HeLa cells were trypsinized, counted, and seeded in 6 cm dishes with 5 mL medium at appropriate concentrations. There were six dishes for each dose. The cells were incubated for 10 days and then stained with crystal violet. The colonies formations were fixed and the surviving fraction was determined by the proportion of seeded cells following irradiation to form colonies relative to untreated cells as described. Colonies with more than 50 cells were counted. The cell survival curve was estimated by a multitarget single-hit model ( $S = 1 - (1 - e^{D/D_0})^N$ ) (L-Q) and then $D_0$ was calculated, where $S$ is the surviving fraction and $D$ is the radiation dose [27]. The sensitization enhancement ratio (SER) was determined by ratio of radiation dose that resulted in 50% survival of the cells.

*2.4. Cell apoptosis*

To measure apoptosis, HeLa cells were incubated for 72 hours at 37 ℃ with 5% $CO_2$. Then 0.1 mM PEG-coated gold NPs with different sizes were added and then radiated by 2 Gy gamma rays. Annexin V binding was performed using an AnnexinV-FITC kit (BD company, US) with the method described by the manufacturer. The cells were washed twice with cold phosphate-buffered saline and were then resuspended in 1× binding buffer at the concentration of $1 \times 10^6$ cells/mL. After this, 100 μL of solution ($1 \times 10^5$ cells) was transferred to a 5 mL culture tube. Annexin V-FITC 5 μL and propidium iodide 5 μL were added, and the cells were then incubated for 15 minutes at room temperature in the dark, after which 400 μL 1× binding buffer was added to each tube and analyzed in the FACS.

### 2.5. Microscopic properties of gold NPs

For the TEM analysis, samples were prepared in dimethyl sulfoxide. The cells were deposited on Formvar-coated 200–300 mesh copper grids and dried, fixed with 2.5% glutaraldehyde in 0.1 M phosphate buffer (pH 7.4), and left in phosphate-buffer for 2 hours. The cells were transferred onto 200-mesh uncoated copper grids, and were observed with a Hitachi HF-2000 field emission high-resolution TEM operating at 200 kV.

### 2.6. Animal model and treatment

All animals were purchased, maintained, and handled using protocols approved by the Institute of Radiation Medicine at the Chinese Academy of Medical Sciences (CAMS). The U14 tumor models were generated by subcutaneous injection of $2 \times 10^6$ cells in 50 μL PBS into the right shoulder of female BALB/c mice. The mice were treated with the PEG-coated gold NPs when the tumor volume reached 50 to 100 mm$^3$ (6 days after tumor inoculation). For the treatment, 100 μL 4.8, 12.1, 27.3, and 46.6 nm PEG-coated gold NPs and saline were injected into mice via intraperitoneal injection. Subsequently, the mice were radiated by 5 Gy gamma rays. 80 mice were assigned to following groups (eight mice in each group): control, 4.8 nm, 12.1 nm, 27.3 nm, 46.6 nm, radiation alone, 4.8 nm + radiation, 12.1 nm + radiation, 27.3 nm + radiation, and 46.6 nm + radiation groups. The injected doses were normalized to be about 4 mg/kg PEG-coated gold NPs. The tumors sizes were measured every two or three days and calculated as: volume = (tumor length) × (tumor width)$^2$/2.

### 2.7. Pathology, immune response, and biochemistry

After 24 days treatment, all mice were sacrificed and blood was collected for biochemistry and organ studies. Mice were sacrificed using isoflurane anesthetic and angiocatheter exsanguination with phosphate-buffered saline. One mouse from each group was fixed with 10% buffered formalin following phosphate-buffered saline exsanguination. During necropsy, liver, kidneys, spleen, heart, and thyroid were collected and weighed. The liver, kidneys, spleen, bone marrow, and lymph nodes were sectioned from the fixed mice. To explicitly examine the grade of changes caused by malities, spleen and thymus indexes ($S_x$) can be defined as:

$$S_x = \frac{\text{Weight of experimental organ } (mg)}{\text{Weight of experimental animal } (g)}$$

Mice were weighed and assessed for behavioral changes. Using a standard saphenous vein blood collection technique, blood was drawn for hematology analysis (potassium EDTA collection tube) and serum biochemistry analysis (lithiumheparin collection tube). Major organs from those mice were harvested, fixed in 4% neutral buffered formalin, processed routinely into paraffin, and stained with hematoxylin and eosin (H&E). Pathology was examined using a digital microscope. Bone marrow and blood cells were obtained after intraperitoneal injection of 24 days at the dose of 4 mg/ml, and were observed with a Hitachi HF-2000 field emission high-resolution TEM operating at 200 kV.

### 2.8. Statistical analyses

All data presented are the average ± SD of experiments repeated three or more times. The paired Student's t-test was used for statistical analysis of luciferase reporter assay and clonogenic assay using Statistical Analysis System (SAS) software.

## 3. Results and discussions

### 3.1. Fabrication and properties of the PEG-coated gold NPs

The size distribution and optical absorption of the PEG-coated gold NPs are characterized by TEM and spectrophotometer in Fig.1. The average sizes of the PEG-coated gold NPs are about 4.8, 12.1, 27.3, and 46.6 nm respectively, which are similar to the previous reported results [1-12]. The surface plasmon resonance (SPR) has been observed for all sizes of the PEG-coated gold NPs by using optical absorption, and they are located at 516, 522, 526, and 535 nm, corresponding to 4.8, 12.1, 27.3, 46.6 nm PEG-coated gold NPs, respectively. It is well known that SPR peak is closely related to medium and physical dimensions, and the increase of size can induce the redshift of SPR. Mie theory can well describe the size-related SPR shift by electromagnetic interaction between gold NPs and incident light [28]. Besides, surface chemical activity can be modified by adding PEG coating. We find that the ζ potentials of 4.8, 12.1, 27.3, and 46.6 nm naked gold particles are -22.1, -18.23, -16.7, and -10.27 mV, respectively, and these high ζ potentials are due to citric acid coating and negative charge in the surface. However, PEG-SH coating induces the decreases of ζ potentials to -6.96, -2.55, -2.097, and -1.065 mV, respectively.

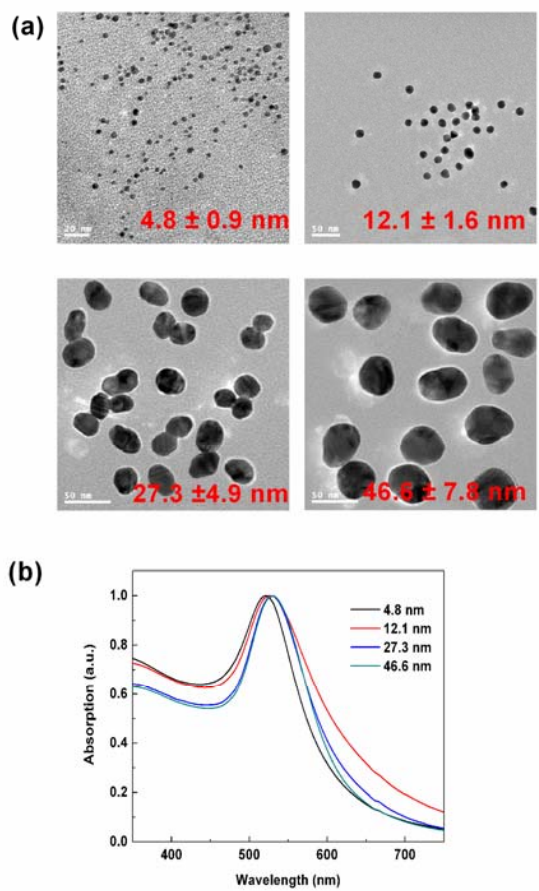

**Figure 1**. (a) TEM image and (b) optical absorption of 4.8, 12.1, 27.3, and 46.6 nm PEG-coated gold NPs.

*3.2. Blood plasma stability of the PEG-coated gold NPs*

To investigate the stability of human blood plasma treated by the PEG-coated gold NPs, the time-dependent SPR peak location and intensity of the PEG-coated gold NPs in human blood plasma have been investigated (Figure 2 (a)). The intensity of SPR peak gradually decreases for all sizes of the PEG-coated gold NPs with increasing time. The decreased intensity of SPR peak is related to the structure and optical properties of the PEG-coated gold NPs, which indicates that the PEG-coated gold NPs combine with the proteins in blood plasma and form larger compounds. A slight redshift about 3-5 nm of SPR peak location is observed in all sizes of the PEG-coated gold NPs. This shift indicates that the PEG-coated gold NPs aggregate slightly in solution, but the SPR peak location has gradually stabilized and has no shift after 6 h. The gradually stabilized SPR peak indicates that the PEG-coated gold NPs can be stabilized. It is well known that the naked and citrate coated gold NPs have high ζ potentials, and they prefer to aggregate in the physiological environment. It has been shown that the naked gold NPs can induce the formation of protein-based aggregation at physiological pH, and this structure is a compound of networks of gold NPs and protein [23, 29]. PEG-coating can improve the surface properties of gold NPs [30]. However, it can be conceived that many serum proteins in the medium can cause the degradation of PEG coating, and form small range aggregation of several particles. In the processing of cellular uptake, the size of the PEG-coated gold NPs can mediate receptor-ligand binding constants, receptor recycling rates, and exocytosis [31]. To directly observe the distribution of the PEG-coated gold NPs in HeLa cell, we have performed TEM observations.

TEM images show microscopic observation into the distribution of 4.8, 12.1, 27.3, and 46.6 nm PEG-coated gold NPs in HeLa cells in Figure 2 (b). After 12 hours cellar uptake, 4.8 nm PEG-coated gold NPs show a bubble aggregation in the cell membrane. Moreover, the PEG-coated gold NPs have local distribution hybrided with biomolecules, which indicates that *in vitro* distribution of 4.8 nm PEG-coated gold NPs is not well monodispersed in the medium and HeLa cell. The complex compounds, including the biomolecules and gold, can be found in the medium. Recently, Oh *et al*. also found that 5.5 nm PEG-coated gold NPs could be observed in the membrane, which is in good agreement with our results [32]. As a contrast, 12.1 nm PEG-coated gold NPs has a wide distribution in intracellular and extracellular areas, which is consistent with the previous results [9]. 12.1 and 27.3 nm PEG-coated gold NPs slightly aggregate by several particles and form larger size of 35-60 nm, which is in agreement with the results of blood plasma stability. 46.6 nm PEG-coated gold NPs can also form larger size hybrid compounds. However, the overall numbers of these particles are very small compared with the data of 12.1 and 27.3 nm particles.

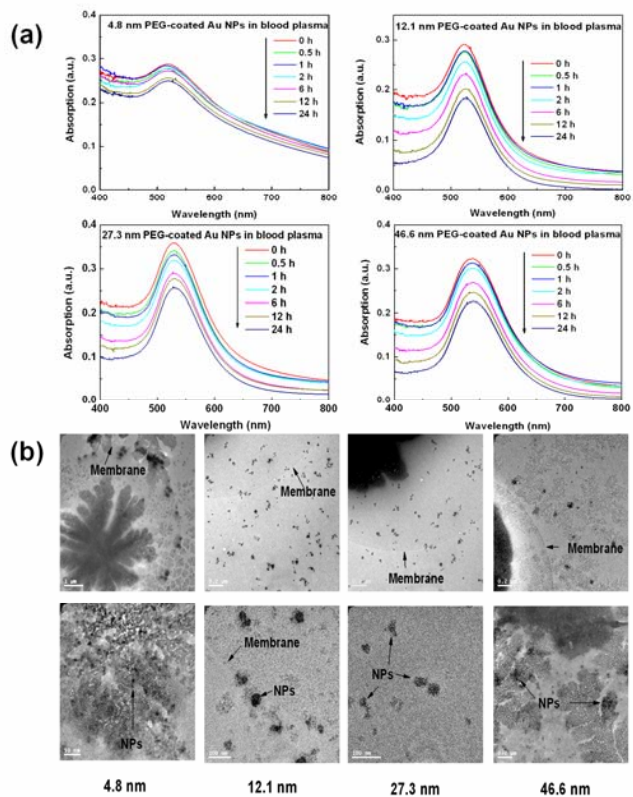

**Figure 2**. Plasma stability of the PEG-coated gold NPs in the human blood plasma have been tested using time-dependent optical absorption. (a)Time-dependent SPR optical absorption and (b) *in vitro* biodistribution of the PEG-coated gold NPs are presented.

*3.3. Size-dependent in vitro cytotoxicity*

The evaluation of cytotoxic effects induced by the PEG-coated gold NPs is performed. After 24 and 48 hours, the viabilities of the cells treated by 4.8, 12.1, 27.3, and 46.6 nm PEG-coated gold NPs are presented at different concentrations of 0.005-0.25 mM in Figures 3 (a) and (b). In general, there is a decrease in cellular viability in a concentration-dependent manner, which indicates that the cytotoxic effects of all sizes of the PEG-coated gold NPs increase in a clear concentration-dependent manner. These data are in agreement with widely reported investigations where the safe dose of the PEG-coated gold NPs was about $10^{-4}$ M [7, 33]. After 24 hour treatments, 4.8 nm PEG-coated gold NPs show the highest toxicity, and cellular viability sharply decreases to 50% above 0.25 mM. The viabilities of the cells treated by 12.1 and 27.3 nm PEG-coated gold NPs also decrease to 60% at the dose of 0.25 mM, but 46.6 nm PEG-coated gold NPs show cellular viability of 90% even at the dose of 0.25 mM. After further treatment of 48 hours, it is found that the viabilities of the cells treated by 12.1 and 27.3 nm PEG-coated gold NPs at 0.25 mM have recovered to 75% and 76%, respectively. However, the cell viability treated by 4.8 nm PEG-coated gold NPs further decreases to 43% at the dose of 0.25 mM, while viability of 83% is observed in cells treated by 46.6 nm PEG-coated gold NPs. By analyzing the dose–response curve, $IC_{50}$ of 4.8, 12.1, 27.3, and 46.6 nm PEG-coated gold NPs after 48 hours are 0.205, 0.477, 0.448, and 0.613 mM, respectively. It can be determined that 46.6 nm PEG-coated gold NPs have the best biocompatibility. 12.1 and 27.3 nm PEG-coated gold NPs have low toxicity while 4.8 nm PEG-coated gold NPs show the highest toxicity. Size-dependent *in vitro* cytotoxicity of the PEG-coated gold NPs has been well documented and it is generally accepted that the smaller a particle is, the greater the toxicity is.

The cytotoxicity of nanomaterials depends on many factors [6]. It is well known that size and surface chemistry of NPs can induce different *in vitro* toxicity effects. It was found that 1.2 and 1.4 nm gold clusters had higher cytotoxicity than 15 nm gold particles by cell necrosis and apoptosis, and modified gold NPs seem closely related to size instead of ligand chemistry [7, 34]. Indeed, 18 nm gold NPs were nontoxic and surface modifiers showed cytotoxicity, while 33 nm gold NPs were found to be low toxicity in different cell lines [34]. This cytotoxicity could be related to the electrostatic adsorption between the cationic NPs and the negatively charged cell membranes. After modification by PEG-SH, gold NPs have lower cytotoxicity and wider cell uptake, which can provide valuable information for further surface modification of therapeutic NPs. For large size of gold NPs, uptake into HeLa cells demonstrats that 46.6 nm PEG-coated gold NPs are nontoxic up to 1 mM. Naked 50 nm gold particles were taken up more quickly by the cells than smaller and larger NPs [31]. Next, we move to *in vitro* and *in vivo* radiotherapy of the PEG-coated gold NPs.

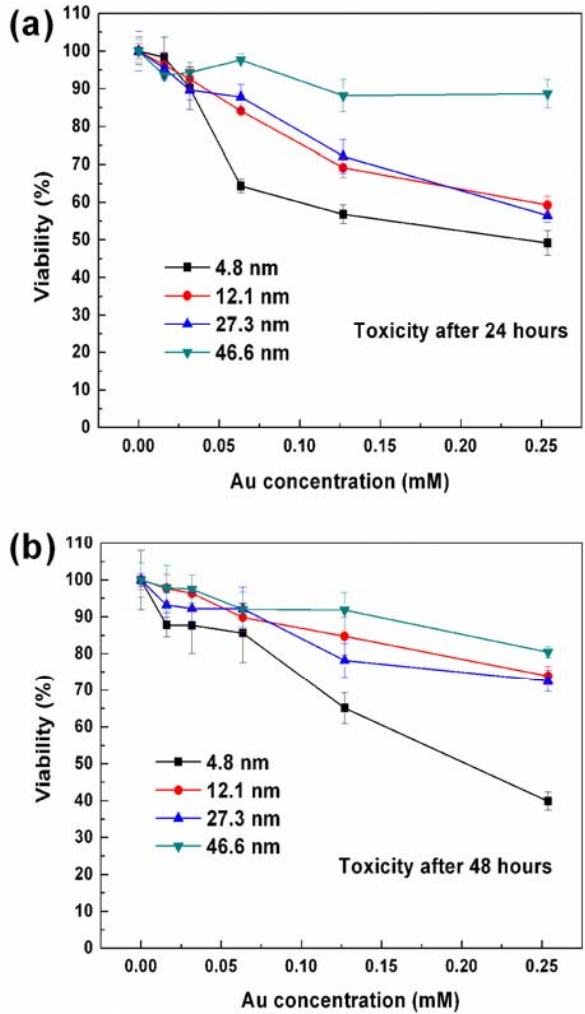

**Figure 3.** Concentration-dependent *in vitro* toxicity of 4.8, 12.1, 27.3, and 46.6 nm PEG-coated gold NPs after (a) 24 and (b) 48 h. The results show that the smaller particles have a higher toxicity.

*3.4. Size-dependent in vitro radiation therapy*

Because gold NPs have shown promise as X-ray contrast agents and radiosensitizers in cancer therapy, we tested *in vitro* radiosensitization effects of the PEG-coated gold NPs in HeLa cells. HeLa cells were treated with PBS control and PEG-coated gold NPs with different sizes at the concentrations of 0.05 and 0.1 mM for 24 hours. Subsequently, the cells were irradiated with 0, 2, 4, 6, and 8 Gy of ionizing radiation, and then rinsed and subjected to clonogenic assays.

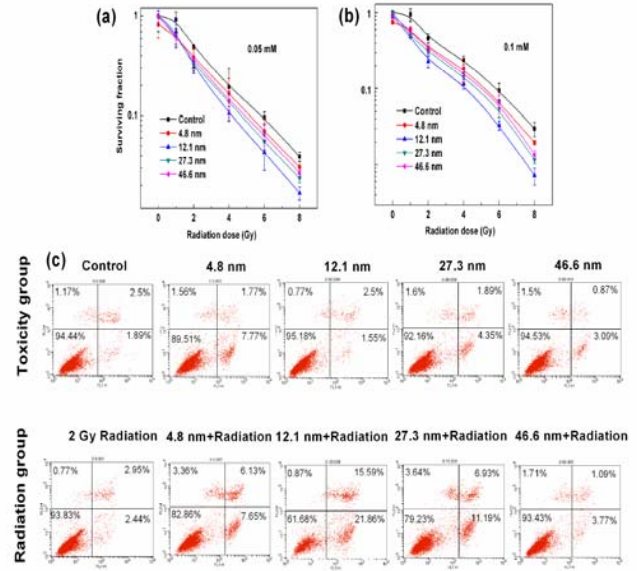

**Figure 4.** Cloning formations of 4.8, 12.1, 27.3, and 46.6 nm PEG-coated gold NPs at the concentrations of (a) 0.05 mM and (b) 0.1 mM using $^{137}$Cs with 662 keV. (c) Cell apoptosis of 4.8, 12.1, 27.3, and 46.6 nm PEG-coated gold NPs after 2 Gy radiation. 12.1 nm PEG-coated gold NPs have the strongest radiation enhancement effect by cellar apoptosis and necrosis.

The radiation dose-dependent radiosensitizing effects of all sizes of the PEG-coated gold NPs (0.05 mM) have been shown in Figure 4 (a). After normalizing survival curves, the surviving fraction decreases, and there is a significant separation of the curves compared with that of radiation alone (P < 0.05). 12.1 nm PEG-coated gold NPs perform the strongest radiation enhancement effect among the four kinds of particles. 27.3 nm PEG-coated gold NPs also shows high enhancement of radiation effect. Furthermore, to evaluate the enhancement efficiency, the SER is calculated. It can be found that the SER of 4.8, 12.1, 27.3, and 46.6 nm PEG-coated gold NPs are 1.41, 1.65, 1.58, and 1.42 respectively, at the concentration of 0.05 mM. It can be found that the size increase of the PEG-coated gold NPs can not increase the radiation sensitivity of HeLa cells. To obtain repeatability, we also carried out experiments to understand the size-dependent radiosensitization effects of the PEG-coated gold NPs at the high concentration of 0.1 mM. As shown in Figure 4 (b), 12.1 and 27.3 nm PEG-coated gold NPs cause higher growth inhibition after 8 Gy radiation than the only radiation group. The calculated SER are 1.46, 2.07, 1.86, and 1.52, corresponding to 4.8, 12.1, 27.3, and 46.6 nm PEG-coated gold NPs, respectively. The results indicate that 12.1 and 27.3 nm PEG-coated gold NPs can effectively sensitize HeLa cells by radiation. To further confirm these results and reveal the radiosensitivity mechanism, the cell apoptosis of the PEG-coated gold NPs was carried out.

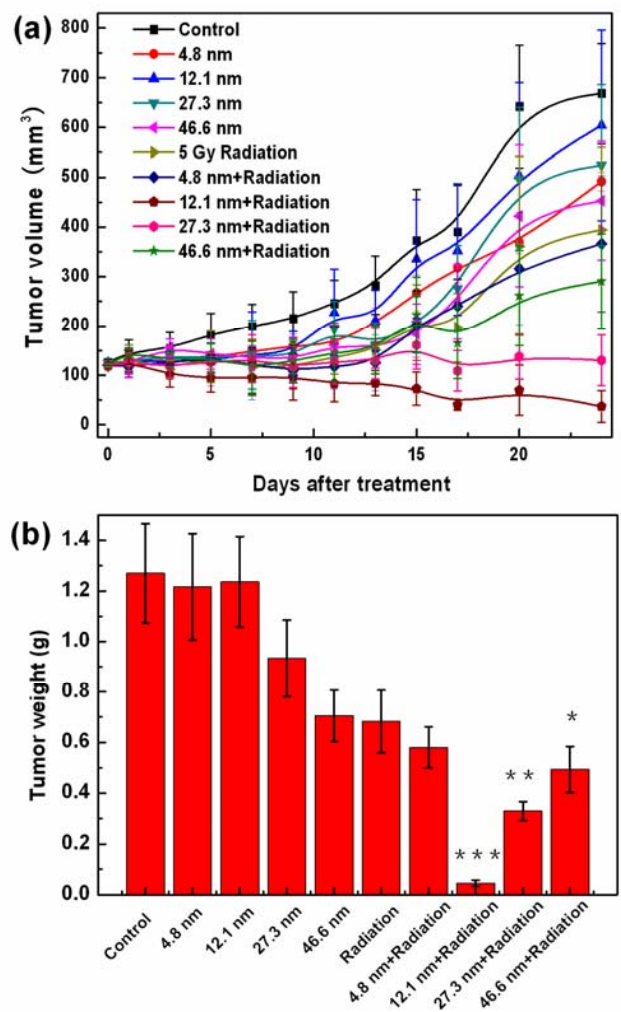

**Figure 5**. (a) Tumor growth curves and (b) tumor weights of 4.8, 12.1, 27.3, and 46.6 nm PEG-coated gold NPs treated mice at the concentration of 4 mg/kg. Bars represent mean ± standard deviation. Data are analyzed by Student's t-test, and *, $P < 0.05$; **, $P < 0.01$; ***, $P < 0.001$. 12.1 nm PEG-coated gold NPs have the most significant decrease of tumor size, and 27.3 nm followed. 4.8 and 46.6 nm PEG-coated gold NPs have the weakest enhancement effects.

The size-dependent cell apoptosis of four kinds of gold NPs at the concentration of 0.1 mM is presented in Figure 4 (c). To estimate whether the PEG-coated gold NPs cause preferentially apoptotic or necrotic cell death, we treat HeLa cells in the logarithmic growth phase with the toxicity groups including control, 4.8, 12.1, 27.3, and 46.6 nm PEG-coated gold NPs. We also treat them with 2 Gy radiation including radiation, 4.8 nm + radiation, 12.1 nm + radiation, 27.3 nm + radiation, and 46.6 nm + radiation PEG-coated gold NPs groups. It can be observed that control group remains 94.44% live, and apoptotic and necrotic cells constitute only 4.39%. For the toxic groups (no-radiation), apoptosis ratios are 7.77%, 1.55%, 5.34%, and 3.09%, corresponding to 4.8, 12.1, 27.3, and 46.6 nm PEG-coated gold NPs, respectively. 4.8 nm PEG-coated gold NPs show slight toxicity by apoptosis, and 12.1, 27.3, and 46.6 nm PEG-coated gold NPs perform tiny cell apoptosis. These results are in agreement with toxicity results. Similar cell apoptosis of 1.4 nm gold clusters was observed, which further confirmed that small gold NPs were more toxic than large gold NPs [34]. To further investigate the radiation sensitivity of PEG-coated gold NPs, apoptosis of cell treated by 4.8, 12.1, 27.3, and 46.6 nm PEG-coated gold NPs + 2 Gy radiation are recorded. Apoptosis ratios induced by radiation are 7.65%, 21.86%, 11.19%, and 3.77%, corresponding to 4.8, 12.1, 27.3, and 46.6 nm particles + 2 Gy radiation treated, respectively. Meanwhile, necrotic cell ratios induced by radiation are 6.13%, 15.59%, 6.93%, and 1.09% for 4.8, 12.1, 27.3, and 46.6 nm particles + 2 Gy radiation treated, respectively. These results indicate that the radiation enhancement effects of 12.1 and 27.3 nm PEG-coated gold NPs can be caused by both rapid cell necrosis and subsequent apoptosis. Meanwhile, the main enhancement effects of 4.8 and 46.6 nm PEG-coated gold NPs are rapid cell necrosis. In general, it can be determined that 12.1 nm PEG-coated gold NPs has the strongest radiation sensitivity among these particles. This can effectively explain the survival fraction from the cloning formation. Indeed, it was found that 10-20 nm PEG-coated gold NPs had a higher efficiency in endocytosis than 45 or 60 nm PEG-caoted gold NPs. Meanwhile, 5 nm gold NPs coated by PEG rarely participated in endocytosis, which was different from the activity of the naked gold NPs [35]. The Xia group also found that the PEG-modified Au nanostructures could hardly adhere to the cell membrane, so their adsorption rate by the cells was very low. Further, the cells had a 1.5–2.4 times higher uptake for the 15 nm NPs than for the 45 nm NPs, which was quite different from the performance of the naked gold particles [36]. Therefore, 12.1 and 27.3 nm particles show higher radiation enhancements than other sizes particles. To test the viability of these results, *in vivo* radiosensitization was carried out.

*3.5. Size-dependent in vivo radiation therapy*

We show *in vivo* cancer treatment by the PEG-coated gold NPs in the U14 mice cancer model in Figure 5. Actually, *in vitro* cultures can't replicate the complexity of an *in vivo* system or provide exact data about the response of a physiologic system to gold NPs. The enhanced *in vivo* radiotherapy is determined by many parameters, including dose, route of exposure, metabolism, excretion, and immune response. Female and male BALB/c mice bearing subcutaneous inoculated U14 tumors were treated by different sizes of the PEG-coated gold NPs over 24 days. The mice were observed daily for clinical symptoms, and the tumor volume was measured by a caliper in tumor-bearing mice. A time-related increase in tumor volume is observed in the control untreated group, 4.8, 12.1, 27.3, and 46.6 nm group, in which the tumors show average volumes of 668.4, 492.3, 604.6, 524.6, and 452.8 mm$^3$, respectively (Figure 5 (a)). The slightly decreased tumor size induced by 4.8, 12.1, 27.3, and 46.6 nm PEG-coated gold NPs can be observed. This represents tumor growth inhibition, but there is no significant statistical difference in inhibition when compared with the control group. After receiving 5 Gy gamma radiation, tumor volumes are significantly decreased. It is found that average tumor volumes are 394.4 (P=0.18 versus control), 349.6 mm$^3$ (P=0.06 versus control), 37.1 (P＜0.001 versus control), 130.3 (P＜0.01 versus control), and 290.4 mm$^3$ (P＜0.05 versus control) for the radiation, 4.8 nm + radiation, 12.1 nm + radiation, 27.3 nm + radiation, and 46.6 nm + radiation PEG-coated gold NPs groups, respectively (24 days). This represents enhanced tumor growth inhibition. 12.1 nm PEG-coated gold NPs show the most significant radiation enhancement effect, and 46.6 nm PEG-coated gold NPs show the lowest enhancement effect. The tumor weights of mice after 24 days are shown in Figure 5 (b). The tumor weights of control, 4.8, 12.1, 27.3, and 46.6 nm PEG-coated gold NPs groups are 1.27, 1.21, 1.24, 0.93, and 0.71 g, respectively. After 5 Gy radiation, the tumor weights decrease to 0.68, 0.53, 0.046 (P＜0.001 versus radiation), 0.33 (P＜0.01 versus radiation), and 0.49 g (P＜0.05 versus radiation), corresponding to radiation, 4.8 nm + radiation, 12.1 nm + radiation, 27.3 nm + radiation, and 46.6 nm + radiation groups, respectively. The tumor weights of mice treated by 12.1 nm PEG-coated gold NPs sharply decrease and nearly disappear after 24 days. Treatment by 27.3 nm PEG-coated gold NPs also leads to a significant decrease. However, 46.6 nm PEG-coated gold NPs has a minimal decrease compared with alone radiation group. These *in vivo* results further confirm that 12.1 nm PEG-coated gold NPs has the greatest radiation enhancement effect.

*3.6. Biodistribution*

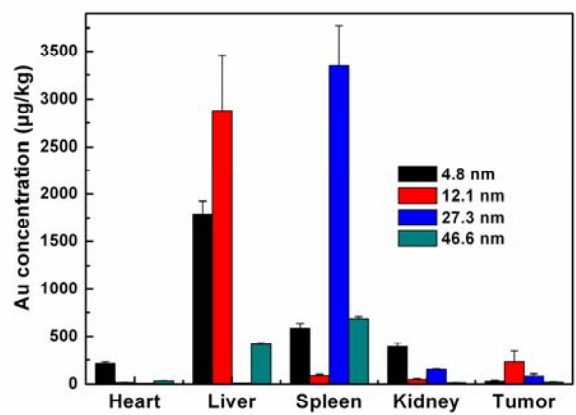

**Figure 6**. Biodistributions of 4.8, 12.1, 27.3, and 46.6 nm PEG-coated gold NPs treated mice after 24 days treatment. 12.1 and 27.3 nm PEG-coated gold NPs have higher content in tumor than that of 4.8 and 46.6 nm PEG-coated gold NPs.

Figure 6 gives biodistributions of the PEG-coated gold NPs in blood and bone marrow cells after intraperitoneal administration at the dose of 4 mg/kg (24 days). Biodistribution shows that liver and spleen are the main target organs, which are in good agreement with the previous results [10, 26, 37]. 4.8 nm PEG-coated gold NPs have a wider distribution in all organs, and 12.1 nm

PEG-coated gold NPs mainly distribute in liver. 27.3 nm NPs prefer to stay in spleen and 46.6 nm NPs have little distribution in all organs. It is notable that 12.1 nm NPs have the highest content in tumor, followed by 27.3 nm NPs. As a contrast, 4.8 and 46.6 nm PEG-coated gold NPs have very tiny accumulation in the tumors. This gives a robust support to *in vivo* therapy results. In available *in vivo* experiments, it was found that 12.1 nm PEG-coated gold particles could rapidly diffuse throughout the tumor matrix, while 45 nm PEG-coated gold particles could only stay near the vasculature [38]. This result indicated long blood circulation time of gold NPs, which was in agreement with the previous reports [37].

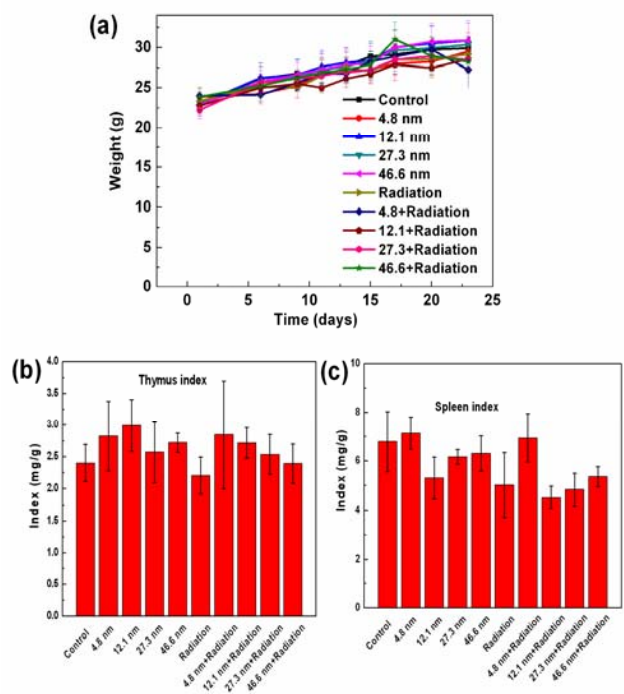

**Figure 7.** (a) Body weight, (b) spleen index, and (c) thymus index of 4.8, 12.1, 27.3, and 46.6 nm PEG-coated gold NPs treated mice at the concentration of 4 mg/kg. Bars represent mean ± standard deviation. Data are analyzed by Student's t-test, and no significant difference from the control group (p < 0.05) is observed.

From the viewpoint of radiation physics, the dominant radiation damage of gold NPs can be caused by photoelectric absorption and secondary electron. The high energy gamma has a strong penetration, and gold NPs of different sizes have nearly similar damage effects and negligible differences of radiation enhancement effect on the HeLa cell. But secondary electron of 0-50 eV can only have a 1-30 nm projected range, which determines the size-dependent protein and DNA damages. Indeed, it was found that gold NPs could enhance degradation of DNA under a high energy ray. There was also a powerful evidence which showed that strand break of DNA could be induced by gold NPs [39]. Besides, distribution and number of the PEG-coated gold NPs in the tumor will play a very important role in the radiosensitization. 12.1 and 27.3 nm PEG-coated gold NPs have wide distributions and an appreciable number in tumor, and cause the massive radiation damage. Therefore, it is not difficult to understand why 12.1 nm PEG-coated gold NPs have the highest enhancement effect, and 4.8 nm PEG-coated gold NPs show negligible radiation enhancement. Increasing size has not induced an enhancement of radiation effect.

*3.7. In vivo toxicity: Pathological, Immune system, and Biochemistry*

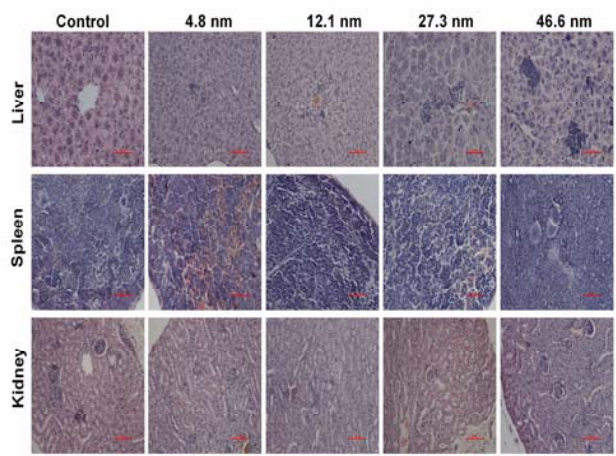

**Figure 8.** Pathological results from the liver, spleen, and kidney of 4.8, 12.1, 27.3, and 46.6 nm PEG-coated gold NPs treated mice at the concentration of 4 mg/kg. All these organs don't show appreciable pathological changes except for liver.

Toxic side effects of gold NPs to normal organs and to overall well-being have been the main problems resulting from cancer radiotherapy and drug delivery. Figure 7 shows (a) body weight, (b) spleen index, and (c) thymus index results. We observe neither considerable body weight loss nor mortality variation in all treated mice (Figure 7 (a)). We perform spleen index (Figure 7 (b)) and thymus index (Figure 7 (c)) to examine immune system damage induced by the PEG-coated gold NPs in mice. The average values of thymus and spleen indices in the control group are 2.39 and 6.8, respectively. High spleen index indicates the immune response of mice, which has been widely reported [40]. After 4.8, 12.1, 27.3, and 46.6 nm PEG-coated gold NPs treated, spleen indices of mice change to 7.1, 4.6, 6.2, and 6.3, respectively. Radiation induces further change to 4.5, 7.1, 4.2, 4.3, and 4.6, corresponding to radiation, 4.8 nm + radiation, 12.1 nm + radiation, 27.3 nm + radiation, and 46.6 nm + radiation group, respectively. Thymus indices of 4.8, 12.1, 27.3, and 46.6 nm PEG-coated gold NPs non-irradiated group increase to 2.8, 2.99, 2.58, and 2.74, respectively. Thymus indices of radiation group, 4.8, 12.1, 27.3, and 46.6 nm PEG-coated gold NPs radiation group are 2.2, 2.8, 2.72, 2.52, and 2.4, respectively. However, it has no significant difference. 12.1 nm PEG-coated gold NPs after 5 Gy radiation show the lowest spleen index, which is consistent with the decrease of tumor volume. Besides, biodistribution of the PEG-coated gold NPs can also influence the organ index. Liver and spleen are considered as two dominant organs for biodistribution and metabolism of the PEG-coated gold NPs. 12.1, 27.3, and 46.6 nm PEG-coated gold NPs have no significant effects on the immune system. Figure 8 presents pathological results for liver, spleen, and kidney. No significant abnormal damages are observed in kidney and spleen. However, liver damage can be seen in mice treated by all sizes of the PEG-coated gold NPs from pathological results.

**Figure 9.** Blood biochemistry results of 4.8, 12.1, 27.3, and 46.6 nm PEG-coated gold NPs treated mice after 24 days. These results show mean and standard deviation of (a) ALT, (b) AST, (c) albumin, and (d) globulin. Data are analyzed by Student's t-test, and the significant difference from the control group ($p < 0.05$) is observed. The increases of ALT and ALS indicate that the liver function has been affected by 12.1, 27.3, and 46.6 nm PEG-coated gold NPs.

Further, a blood chemistry test was performed after 24 days of initiation treatment. Specific attention was paid to those hepatic-related serum chemistries that would reflect liver damage or alternation of liver function. It is found that 12.1, 27.3, and 46.6 nm PEG-coated gold NPs, including radiation and non-radiation, can cause the increases of aspartate aminotransferase (AST) and alanine transaminase (ALT) in Figure 9. For alone radiation group, ALT also increases. Besides, albumin (ALB) and globulin (GLOB) are found to have no significant statistical difference. Indeed, it is not difficult to understand this phenomenon. 13 nm gold particles coated by PEG could induce abnormal gene expression and accumulation in liver [8, 10, 41]. In spite of reducing the toxicity of gold NPs by PEG coating, the biodistribution of 13 nm PEG-coated gold particles showed long retention in liver [8, 10]. These results showed that further metabolism of gold based materials must be investigated. *In vivo* toxicity and biodistribution show that the PEG-coated gold NPs are different from the naked gold NPs. 12.1 nm PEG-coated gold NPs have a higher efficient diffusion and absorption for tumors and can further mediate tumor targeting efficiency. Naked gold NPs strongly associated with these essential blood proteins (such as albumin, fibrinogen, γ-globulin, histone, and insulin), and the degree of binding between NPs and proteins was related to the NPs size [29, 42]. Furthermore, these interactions, such as enhanced binding and whether they can induce conformational change of the proteins, should be further investigated. It would be interesting to fabricate the tumor targeting bioconjunction based on 12.1 and 27.3 nm PEG-coated gold NPs, and to investigate *in vitro* and in *vivo* targeting radiotherapy. Meanwhile, whether these therapeutic particles in liver or kidney can be metabolized by the bioactive environment if the therapy time is above 6 mouths is a very important question for future clinical therapy. It is desirable to find an available route to improve metabolism of the PEG-coated gold NPs. Most recently, we have found the GSH-protected gold nanoclusters perform high-efficient renal clearance, and the enhancement of radiation therapy of these small nanoclusters will be very interesting [43].

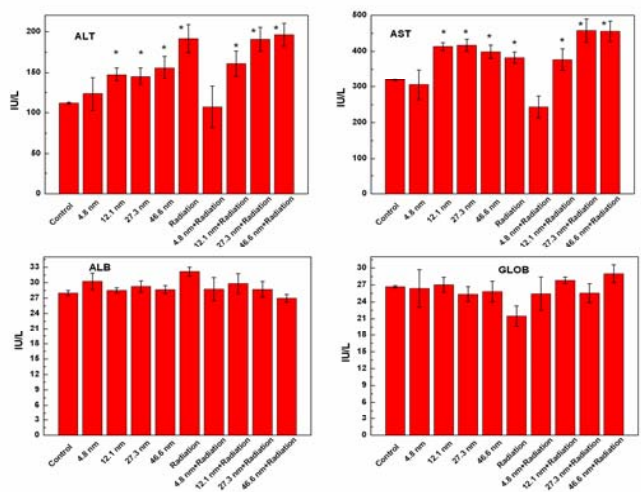

## 4. Conclusion

Size-dependent *in vitro* and *in vivo* radiosensitizations of the PEG-coated gold NPs were carried out. Toxicity, survival fraction, apoptosis, and morphology of HeLa cell were investigated at different concentrations. *In vitro* toxicity results showed that small NPs had higher toxicity than that of large NPs. Further, 12.1 and 27.3 nm PEG-coated NPs had wider distributions in the cells and performed higher radiation enhancement effects than that of 4.8 and 46.6 nm NPs. It was found that 4.8 and 46.6 nm PEG-coated gold NPs caused predominantly rapid cell death by apoptosis, while 12.1 and 27.3 nm PEG-coated gold NPs induced enhancement of radiotherapy by both necrosis and apoptosis. *In vivo* cancer treatments also showed that all sizes of PEG-coated gold NPs could decrease the tumor volume and weight after 5 Gy radiation, but 12.1 and 27.3 nm PEG-coated gold NPs could induce appreciable decrease of tumor volume and weight. Immune response and blood biochemistry from *in vivo* toxicity indicated that the PEG-coated gold NPs did not cause appreciable toxicity except for the slight liver damage. Based on *in vitro* and *in vivo* radiation therapy, it can be conceived that 12.1 and 27.3 nm PEG-coated gold NPs show high radiosensitivity and thus are valuable for possible applications in enhancement of radiotherapy and drug delivery.

ACKNOWLEDGMENTS

This work is supported by the National Natural Science Foundation of China (Grant No.81000668), and the Subject Development Foundation of Institute of Radiation Medicine, CAMS (Grant No.SF1207).